\documentclass[aps,showpacs,12pt]{revtex4}%
\usepackage{graphics}
\usepackage{graphicx}
\usepackage{epsfig}
\usepackage{amsmath}
\usepackage{amsfonts}
\usepackage{color}
\usepackage{amssymb}%
\newcommand {\ber}{\begin{eqnarray}}
\newcommand {\eer}{\end{eqnarray}}
\newcommand{\be}{\begin{equation}}
\newcommand{\ee}{\end{equation}}
\newcommand{\bel}[1]{\begin{equation}\label{#1}}
\newcommand{\bea}{\begin{eqnarray}}
\newcommand{\eea}{\end{eqnarray}}
\newcommand{\ba}{\begin{array}}
\newcommand{\ea}{\end{array}}

\newcommand{\rme}{\mbox{\rm e}}

\DeclareMathOperator{\LEFT}{LEFT}
\DeclareMathOperator{\RIGHT}{RIGHT}
\DeclareMathOperator{\eff}{eff}

\begin{document}
\title{ASEP on a ring conditioned on enhanced flux}
\author{Vladislav Popkov}
\email{popkov@uni-bonn.de}
\affiliation{Interdisziplin\"ares Zentrum f\"ur Komplexe Systeme, Universit\"at Bonn,
R\"omerstr. 164, 53115 Bonn, Germany}
\affiliation{Universit\`a di Salerno, Dipartimento di Fisica "E.R. Caianiello",  via Ponte don Melillo, 84084 Fisciano (SA), Italy}

\author{Gunter M. Sch{\"u}tz}
\email{g.schuetz@fz-juelich.de}
\affiliation{Institut f\"ur Festk\"orperforschung, Forschungszentrum J\"ulich, 52425
J\"ulich, Germany}
\affiliation{Interdisziplin\"ares Zentrum f\"ur Komplexe Systeme, Universit\"at Bonn,
R\"omerstr. 164, 53115 Bonn, Germany}

\author{Damien Simon}
\email{damien.simon@upmc.fr}
\affiliation{Laboratoire "Probabilit\'es et Mod\`eles Al\'eatoires", UPMC, 4 place Jussieu, 75252 Paris Cedex 05, France}
\date{\today}

\begin{abstract}
We show that in the asymmetric simple exclusion process (ASEP) on a ring,
conditioned on carrying a large flux, the particle experience an effective
long-range potential which in the limit of very large flux takes the simple
form $U= -2\sum_{i\neq j}\log|\sin\pi(n_{i}/L-n_{j}/L)|$, where $n_{1}%
,n_{2},\ldots n_{N}$ are the particle positions, similar to the effective potential between 
the eigenvalues of the circular unitary ensemble in random matrices. Effective hopping rates and
various quasistationary probabilities under such a conditioning are found
analytically using the Bethe ansatz and determinantal free fermion techniques. Our
asymptotic results extend to the limit of large current and large activity for a
family of reaction-diffusion processes with on-site exclusion between
particles. We point out an intriguing generic relation between classical stationary
probability distributions for conditioned dynamics and quantum ground state
wave functions, in particular,  in the case of exclusion processes, for free fermions.

\end{abstract}

\pacs{05.40.-a, 02.50.Ey, 82.20.-w}
\maketitle

\section{Introduction}

In this work we investigate the spatio-temporal structure of a family of stochastic
interacting particle systems, specifically exclusion processes, in the case of
the extreme event in which the system supports a very large current for an
extended period of time, i.e., while the system is in a state very far from typical
behaviour. As this type of non-equilibrium behaviour does not have a long history
of study in statistical mechanics, we first make some brief remarks on how this problem is
situated in general terms in the study of the statistical properties of extreme events. Then
we state the precise problem that we are going to address.

When studying statistical properties of extreme events \cite{Albe06}, i.e., of highly 
non-typical values of
some physical observable in a many-body system with noisy dynamics, the first
question to ask is the full probability distribution of that observable since
one wishes to know not only fluctuations around its mean, but also the
distribution very far from it. From that knowledge one may be able to infer
knowledge on the extreme value statistics for that observable. Then, a
second problem of interest is the spatio-temporal patterns
that generate such atypical values of observables over some period of time.
Even though very difficult to tackle, this is an important question since a
good understanding of this aspect of non-equilibrium behaviour, which lies beyond a purely
statistical description in terms of extreme value statistics, may give insight
into the physical mechanisms that generate extreme events.
One could potentially identify and apply forces to enhance the
probability of such rare events. Notice that in this setting we are not interested in searching for
just \textit{any} force that would make some rare large fluctuation of a
physical observable typical but for those very specific forces that retain the
spatio-temporal patterns of the original unforced process that generates these
rare fluctuations by its own intrinsic random dynamics. Conversely, if suitable
forces would suppress the relevant space-time patterns, one could make such
events even more unlikely.

Carrying out such an analysis is obviously very difficult even for relatively simple
complex systems. However, this problem turns out to be interesting already in the
context of simple exactly solvable statistical mechanics models.
To be specific, we consider in this work the Asymmetric Simple
Exclusion Process (ASEP) \cite{Ligg99,Schu01}
and we look for the spatio-temporal patterns
that generate an atypically large flux over a long period of time and for which
effective hopping rates (arising from some extrinsic applied force) would
generate such a state as typical stationary state. Earlier exact work on
currents below the typical value have revealed that long-lasting low-current events are
supported by travelling waves which represent density profiles with two
regions of constant densities $\rho_{1,2}$ \cite{Bodi05}. Such spontaneous
phase separation behaviour has earlier been suggested by Fogedby
\cite{Foge98,Foge02} who used renormalization group arguments to show that
atypical low-current fluctuations are realized by a gas of shock and antishock
solutions of the noisy Burgers equation that describes the large scale
behaviour of the ASEP.  In the context of phase separation these shocks and antishocks
are the domain walls that separate regions of high and low density. On a
microscopic level the shocks are known to be sharp, the bulk properties of
each phase are reached quickly within a small distance from the shock position \cite{Derr97}.

Understanding how the particle system
organizes itself microscopically for atypically {\it large} currents, and how such a microscopic
structure could be generated as typical event by introducing suitable forces,
requires a different and careful exact analysis. A step in this direction was done in
recent work by Simon \cite{Simo09}, where the effective hopping rates of the particles in 
specific regimes of the ASEP with open boundaries were computed. To make further
progress and to investigate an arbitrary number of particles we consider in the present
work periodic boundary conditions. Using Bethe ansatz we compute the
quasi-stationary distribution of the system conditioned on sustaining a large
current and we obtain an exact effective interaction potential between
particles that would make this quasi-stationary distribution stationary, i.e. typical. We also 
stress an interesting analogy with random matrix ensembles.

It will transpire that for very large currents our results for the ASEP are rather generic.
The results remain valid for all
exclusion processes with nearest neighbour hopping, irrespective of the
hopping asymmetry or even particle number conservation. Hence our approach
covers also e.g. reaction-diffusion processes of the type studied in
\cite{Alca93,Schu95}. These include the well-known contact process, diffusion-limited annihilation,
the coagulation/decoagulation model and other well-studied processes.
Moreover, the same results are valid for the limit of large
hopping activity, irrespective of direction. We also point out an intriguing
connection to quantum mechanical expectation values for free fermion systems.

The paper is organized as follows: In the following section (\ref{Approach}) we define the 
problem and
present our approach.  Mainly for pedagogical reasons this is briefly illustrated  for the case of
just two particles (Sec.  \ref{Sec:Twoparticles}) and then in detail for three particles in
Sec.  \ref{Sec:Threeparticles}. From these preliminaries we proceed to the investigation
of an arbitrary number of particles in the limit of very large current in Sec. \ref{Sec:Nparticles}
where our main results are derived. We finish with a brief summary and discuss generalizations
of our results to other particle systems (Sec. \ref{Conclusions}).

\section{ASEP conditioned on carrying an atypical current}
\label{Approach}

In the asymmetric simple exclusion process (ASEP)
\cite{Ligg99,Schu01} in one dimension with periodic boundary
conditions and $L$ sites, particles jump independently after an
exponentially distributed random time with mean $1/(p+q)$ to a
nearest neighbor site, provided that the target site is empty. The
probability to choose the right neighbour (clockwise in a periodic
chain) is $p/(p+q)$ while the probability of choosing the left
neighbour (anticlockwise) is $q/(p+q)$.
Throughout the paper, we assume $p+q=1$ for
notational simplicity.
 If the target site is
occupied the jump attempt is rejected. This exclusion principle
guarantees that each site is always occupied by at most one
particle. An instantaneous configuration of this system with $N$
particles can therefore be
represented by an ordered set of integer coordinates $\{\mathbf{n}%
\}=\{n_{1},\dots,n_{N}\}$ where $n_{k+1}>n_{k}$ and $n_{k} \in
\{1,\dots,L\}$ for all $k$.

Mathematically, the dynamics of the ASEP can be defined through a master
equation for the probability $P_\mathbf{m}(\mathbf{n},t)$ to find a configuration
$\mathbf{n}$ at time $t$ starting from a configuration $\mathbf{m}$ at time $0$. 
The master equation takes the form
\begin{equation}
\frac{d}{dt}P_\mathbf{m}(\mathbf{n},t)=\sum_{\mathbf{n}^{\prime}\neq\mathbf{n}}\left[
w_{\mathbf{n},\mathbf{n}^{\prime}}P_\mathbf{m}(\mathbf{n}^{\prime},t)-w_{\mathbf{n}%
^{\prime},\mathbf{n}}P_\mathbf{m}(\mathbf{n},t)\right]  , \label{2-1}%
\end{equation}
where $w_{\mathbf{n}^{\prime},\mathbf{n}}$ is the transition rate (0, $p$ or
$q$ for the ASEP) to go from a configuration ${\mathbf{n}}$ to a configuration
${\mathbf{n}}^{^{\prime}}$. By integrating (\ref{2-1}) one obtains the
solution of the master equation for any given initial configuration
$\mathbf{m}$, i.e., the conditional probability to find a particle
configuration $\mathbf{n}$ at time $t$, given that the process started from
configuration $\mathbf{m}$. In the stationary distribution of a system with
$N$ particles, all particle configurations are equally likely. Therefore, up to
finite-size corrections of order $1/L$, the stationary particle current has
the simple form $j=(p-q)\rho(1-\rho)$ for particle density $\rho=N/L$. For
details, see e.g. \cite{Ligg99,Schu01}. Since the master equation can be used
generally for describing Markov processes we shall use at some places below
the generic symbol $C$ for microscopic configuration of a process.

From the linear form of (\ref{2-1}) it is obvious that the master equation can
be cast in matrix form
\begin{equation}
\label{2-2}\frac{d}{dt} \mbox{$|  {P(t)} \rangle$} = - H
\mbox{$|  {P(t)}
\rangle$}
\end{equation}
where the probability vector $\mbox{$|  {P(t)} \rangle$}$ has all the
probabilities $P(\mathbf{n},t)$ as its canonical components. In a conveniently chosen
tensor basis the stochastic generator $H$ of the ASEP, often called ``quantum
Hamiltonian'', takes the form \cite{Schu01}
\begin{equation}
\label{2-3}H = - \sum_{k=1}^{L} \left[  p(s_{k}^{+} s_{k+1}^{-} - n_{k}
(1-n_{k+1})) + q(s_{k}^{-} s_{k+1}^{+} - (1-n_{k})n_{k+1})\right] =W_+ + W_- + W_0
\end{equation}
Here $s_{k}^{\pm}= (\sigma_{k}^{x} \pm\sigma_{k}^{y})/2$ are the
$SU(2)$ spin-1/2 ladder operators acting on site $k$ of the
lattice and $n_{k} = (1-\sigma_{k}^{z})/2$ is the projection
operator on states with a particle on site $k$.
The operators $W_+$, $W_-$, and $W_0$ correspond respectively to the pure right-jump 
generator, to the pure left-jump generator and the diagonal part (see below for a detailed
discussion). The stationary distribution is encoded in the right eigenvector of $H$ corresponding 
to the lowest eigenvalue 0. 

The process as defined above describes only the evolution of the microscopic
particle configuration, but does not keep track of the current that flows
during the evolution. This can be achieved by introducing another random
variable which counts the number of jumps across a given bond $k,k+1$ of the
lattice, see \cite{Harr07} for a detailed description. Then each time a
particle jumps across that bond to the right (left) the value of the current
counter is incremented (decremented) by one unit. After some time $t$ this
random variable then provides the integrated current, i.e., the total net
number of jumps $J_{k} (t)$ across bond site $(k,k+1)$ up to time $t$. Because
of particle number conservation one may equivalently one may also count the
total net number of jumps in the lattice. We denote this quantity by $J(t)$.
By the law of large number one expects an asymptotically linear growth $J(t)
\propto jt$ where $j$ is the stationary current of the ASEP.

With these definitions at hand we can sharpen the question of space-time
realizations for extreme currents and define precisely our problem. Following
\cite{Simo09} we consider histories such that for a given initial
configuration a certain integrated current $J$ has flown until some large time
$T$. We then ask how the system behaves during a time interval $[t_{1},t_{2}]$
such that both $t_{1}$ and $t_{2}$ are far from the initial time 0 and the
final time $T$. As shown in \cite{Simo09} this is tantamount to computing an
effective stochastic process that has $j = J/T$ as its typical current and
computing the stationary distribution of that effective process. This yields
the quasi stationary distribution of the original ASEP conditioned on
sustaining an atypical average flux $j$. Before proceeding with the
discussion of this distribution we make two short digressions, the
significance of which will become clear immediately further below.

First we point out that generally the generator of a stochastic process can be
written
\begin{equation}
\label{2-4}H = A - D 
\end{equation}
where matrix elements $A_{C^{\prime},C} = - w_{C^{\prime},C} \leq0$ of the
off-diagonal matrix $A=W_+ + W_-$ is given by the negative transition rates
$w_{C^{\prime},C}\geq0$ from a configuration $C$ to another configuration
$C^{\prime}$, while the diagonal matrix $D=W_0$ is given by the negative sum of
rates out of a given configuration, i.e., $D_{C,C} = - \sum_{C^{\prime}}
w_{C^{\prime},C}$. This property ensures that all diagonal elements of $H$ are
positive real numbers, while all off-diagonal elements are negative real
numbers. Conservation of probability is encoded in the fact that by
construction the matrix elements in each column of $H$ sum up to 0. From this
property it follows that there is a left eigenvector $\mbox{$\langle \,
{s}\, |$}$ with eigenvalue 0 where all components are equal to 1. This
property along with the positivity (negativity) and reality condition on the
matrix elements guarantee that $H$ is the generator of some Markovian
stochastic dynamics.

Generalizing an observation made in \cite{Simo09} we now consider an irreducible matrix
$M$ with non-negative diagonal and non-positive off-diagonal part. By the
Perron-Frobenius theorem its lowest eigenvalue, denoted $\mu$ below, is real
and nondegenerate, and the respective left and right eigenvectors
$\mbox{$\langle \, {\mu}\, |$},\,\mbox{$| \, {\mu}\, \rangle$}$ have all
positive real components $\mu_{C}^{R}=\langle C\mbox{$| \, {\mu}\, \rangle$}$,
$\mu_{C}^{L}=\mbox{$\langle \, {\mu}\, |$}C\rangle$. Let us define a diagonal
matrix $\Delta$ with the components $\mu_{C}^{L}$ on the diagonal and consider
the transformed and shifted matrix
\begin{equation}
M^{\prime}=\Delta M\Delta^{-1}-\mu. \label{2-5}%
\end{equation}
It is easy to see that $M^{\prime}$ has an eigenvalue $0$ with a right
eigenvector $\mbox{$| \, {\mu'}\, \rangle$}=\Delta\mbox{$| \, {\mu}\,
\rangle$}$ and left eigenvector $\mbox{$\, \langle{s}\, | $}=\mbox{$\langle
\, {\mu}\, |$}\Delta^{-1}$ with constant components equal to one. Moreover, by
construction all off-diagonal elements of $M^{\prime}$ are non-positive real
numbers, while the diagonal of $M$ remains unchanged up the constant
real-valued shift $\mu$. Hence the matrix $M^{\prime}$ is the generator of a
Markov process. The construction (\ref{2-5}) thus provides a general recipe
how to obtain a stochastic generator from a rather general family of
non-degenerate matrices with non-positive real off-diagonal and non-negative
real diagonal part. The stationary distribution can be written in terms of the
lowest eigenvectors of $M$ as
\begin{equation}
P_{C}^{\ast}=\mu_{C}^{L}\mu_{C}^{R}/Z \label{2-6}%
\end{equation}
where $Z=\sum_{C}\mu_{C}^{L}\mu_{C}^{R}=\mbox{$\langle \, {\mu}\, |$}\,\mu
\,\rangle$ is the normalization that ensures that the sum of all probabilities
is equal to 1. We remark that for symmetric $M$ one has $\mu_{C}^{L}=\mu
_{C}^{R}$. Then $M$ is also hermitian and therefore can be seen as the
Hamiltonian of some quantum system. We find it intriguing that in such a case
the stationary distribution of the associated classic stochastic dynamics
defined by the Markov generator $M^{\prime}$ is given by the quantum
mechanical, i.e., quadratic probability distribution of the quantum system. 
The classical interpretation of the presence of both the left and right eigenvectors 
in this formula is the following~: conditioning on an atypical current up to a final time 
$T$ requires to produce this current both between $0$ and $t$ and between $t$ 
and $T-t$. Computations made in \cite{Simo09} show that the left eigenvector 
take into account the first part $[0,t]$ whereas the right one takes into account 
the second part $[t,T-t]$; there are both a forward-in-time and a 
backward-in-time effect of the global conditioning.

Next we comment specifically on the current distribution $P(J) = \mbox{\rm
Prob}[J(t)=J]$ in the ASEP. We introduce the time-averaged current $j=J/t$, sometimes
simply called current, as opposed to the integrated current $J$.
Asymptotically the current distribution has the large deviation form
\begin{equation}
\label{2-7}P(J) \propto\mbox{\rm e}^{t f(j)}%
\end{equation}
where $f(j)$ is the large deviation function. Introducing the auxiliary
parameter $s$ conjugate to the integrated current one obtains the scaled
asymptotic cumulant function
\begin{equation}
\label{2-8}\mu(s) = \lim_{t\to\infty} \frac{\ln\langle \rme^{sJ(t)}\rangle}{t} =
\lim_{t\to\infty} \frac{\ln\sum_{J} P(J) \rme^{sJ(t)}}{t}%
\end{equation}
which is given by the largest eigenvalue of the modified rate
matrix\cite{Harr07,DerridaAppert99}
\begin{equation}
\label{ModifiedRateMatrix}\widehat{W}(s)=-(\rme^{s}W_{+} +\rme^{-s}W_{-} +W_{0}).
\end{equation}
Here $W_{+} = - p\sum_{k} s_{k}^{+} s_{k+1}^{-}$ and $W_{-} = - q\sum_{k}
s_{k}^{-} s_{k+1}^{+}$ are the off-diagonal part of the generator of the ASEP,
corresponding to moves to the right and left resp., and $W_{0} = \sum_{k}[ p
n_{k}(1-n_{k+1}) + q (1-n_{k})n_{k+1}]$ is the diagonal part. Thus one obtains
the modified rate matrix from the original generator $H$ (\ref{2-3}) simply by
multiplying all off-diagonal elements with $e^{\pm s}$. Notice the sign
convention that we adopt here to remain consistent with \cite{Simo09}: For
$s=0$ we have $\widehat{W}(0)=-H$.

The two large deviation functions $f(j)$ and $\mu(s)$ are related by a
Legendre transformation
\begin{equation}
\mu(s)=\max_{j}[f(j)+sj]. \label{2-9}%
\end{equation}
This construction is in complete analogy to a change between ensembles in
equilibrium thermodynamics where $J$ would be regarded as extensive variable
(here: extensive in time) whereas $s$ would be the conjugate intensive
quantity. On the one hand one might want to study for a given time $t$
space-time histories with a fixed value of $J=j_{cond}t$. Alternatively, one
could study the transformed ensemble where $J$ is fluctuating, and the
conjugate quantity $s$ is chosen fixed such that the current has the same mean
value
\begin{equation}
j_{cond}(s)=\frac{d}{ds}\mu(s), \label{2-10}%
\end{equation}
By construction this \textquotedblleft conditional current\textquotedblright%
\ is a monotonically increasing function of $s$ and it coincides at the point
$s=0$ with the stationary current of the ASEP. So positive (negative) values
of $s$ correspond to an atypical current enhanced (reduced ) with respect to
the typical stationary current of the ASEP. Equivalently one may write
\begin{equation}
s(j)=-\frac{d}{dj}f(j) \label{2-11}%
\end{equation}
to obtain the inverse relation between $s$ and $j$.

With these notions we can address the question of how the ASEP, conditioned to
produce an atypical current $j$ during a long period of time, behaves
dynamically. We consider this question in the ensemble of fixed $s$ rather
than fixed $j$ where for a given $j$ the conjugate value of $s$ is given by
(\ref{2-11}). As shown in \cite{Simo09} the general construction presented in
our digression can be used to solve this problem. The rates of the effective
Markov process are expressed in terms of the left eigenvector of the modified
rate matrix (\ref{ModifiedRateMatrix}) corresponding to its largest eigenvalue
as
\begin{equation}
W_{C^{\prime}C}^{\eff}=\widehat{W}(s)_{C^{\prime}C}\frac{\langle\mu
_{1}(s)|C^{\prime}\rangle}{\langle\mu_{1}(s)|C\rangle},\text{ for }C\neq
C^{\prime} \label{EffectiveDynamicsRate}%
\end{equation}
Here $W_{C^{\prime}C}$ , $W_{C^{\prime}C}^{\eff}$ are rates of change
$C\rightarrow C^{\prime}$ in the original and in the effective stochastic
process. The unnormalized stationary state vector of the effective process is
given by (\ref{2-6}) and can be written in bra-ket notation as
\begin{equation}
P_{stat}^{\eff}(C)=\langle\mu(s)|C\rangle\langle C|\mu(s)\rangle.
\label{EffectiveDynamicsStationaryState}%
\end{equation}
The stationary flux (\ref{2-10}) of the effective stochastic process is equal
to the atypical flux, on which the original stochastic process is conditioned.

From a probabilistic point of view, the transformation from the initial Markov 
process to the conditioned one with the modified matrix $M'$ correspond 
to a change of measure. The expectation value under the new process of 
any observable $A_t$ at time $t$ can be expressed as:
\begin{equation}
\widehat{E}(A_t) = \lim_{T\to\infty}\frac{E( A_t \rme^{sJ(T)})}{E(\rme^{sJ(T)})}
\end{equation}
The existence of the limit and its relation to the conditioning under the current are 
deduced from the large deviation principle satisfied by the current $J(T)$. Such a 
change of measure makes explicit the conjugation relation between the 
parameter $s$ and the current $J(t)$, similar to equilibrium statistical mechanics.

So, if we succeed to solve the largest eigenvalue problem for the modified
rate matrix (\ref{ModifiedRateMatrix}), the effective stochastic process and
its respective stationary state can be constructed. The aim of this paper is
to construct such an effective stochastic process for the ASEP on a ring,
conditioned to produce a high particle current. The stationary distribution of
the unconditioned ASEP does not have long-range correlations and the process
itself has only has hard-core on-site interaction. Conditioning the process on
an atypical high flux, one expects that the respective space-time particle
trajectories (for long periods when this high flux can be observed), avoid
forming clusters, thus generating an effective repulsive interaction. In this
paper we find the analytical form of such interaction for very high fluxes.
The ASEP is chosen for simplicity of presentation, but it will transpire that
our results are generalizable to non-conservative simple exclusion processes
with nearest neighbour hopping, i.e. to a class of reaction-diffusion processes.

We consider first the special cases of two and three particles on a ring of
$L$ sites, and then generalize to the case of an arbitrary number of particles
$N$. For calculations we use the fact that the modified rate matrix for
the ASEP belongs for all $s$ to the class of integrable models whose
eigenfunctions can be found using the Bethe Ansatz.

We stress that three different connections of classical stochastic processes
to quantum mechanics are described
in this work. They are independent from each other and have their own meaning. 
The first one presented in the previous section is the quantum Hamiltonian formalism 
to describe the evolution of the probability distribution. It arises from the Markov
property of the stochastic process, in particular from the linearity of the temporal evolution
of the process. However, the classical probabilities are linear in the right eigenvector,
while quantum mechanical expectations are quadratic. The second connection, which is
{\it new}, stems from the effective dynamics where the conditioned expectation values are 
now quadratic in the eigenvectors (see \eqref{EffectiveDynamicsStationaryState}) as
in quantum expectations. This fact is independent of the previous quantum formalism. 
Finally, the determinantal free fermion structure described in the next sections for a large 
current is also a \emph{new} feature of the present regime of the exclusion process and 
is not a direct consequence of the two previous facts.

\section{Two particles on a ring}
\label{Sec:Twoparticles}

The eigenfunction of the modified rate matrix $\widehat{W}$ is found using the
coordinate Bethe ansatz (for the present context see e.g.
\cite{Kim95,Schu97,DerridaAppert99})
\begin{equation}
\mbox{$| \, {\varphi}\, \rangle$}=\sum\limits_{1\leq n1<n_{2}\leq L}\left(
z_{1}^{n_{1}}z_{2}^{n_{2}}+A_{21}z_{2}^{n_{1}}z_{1}^{n_{2}}\right)
|n_{1}n_{2}\rangle=\sum\limits_{1\leq n1<n_{2}\leq L}Y_{n_{1}n_{2}}|n_{1}%
n_{2}\rangle
\end{equation}
This eigenvector has eigenvalues $\Lambda=p\rme^{s}/z_{1}+p\rme^{s}/z_{2}%
+q\rme^{-s}z_{1}+q\rme^{-s}z_{2}-2$ and the amplitude $A_{21}$ is given by
requirement that $Y_{n_{1}n_{2}}$ is an eigenfunction for particles being
nearest neighbours
\begin{equation}
A_{21}=-\frac{p\rme^{s}+q\rme^{-s}z_{1}z_{2}-z_{2}}{p\rme^{s}+q\rme^{-s}z_{1}z_{2}-z_{1}}.
\label{A21}%
\end{equation}
Moreover, periodic boundary conditions require
\begin{equation}
z_{2}^{L}=A_{21}. \label{A21per}%
\end{equation}

The translation invariance of the problem together with the constant sign of 
the components of the ground state yields $Y_{n_{1}n_{2}%
}=Y_{n_{1}+1,n_{2}+1}$ and therefore $z_{1}z_{2}=1$ for the ground state. It is convenient to
parametrize $z_{2}=e^{i\gamma},\,z_{1}=e^{-i\gamma}$. Then (\ref{A21per}) is a
transcendental equation which has $L$ solutions for $\gamma$. The amplitude
corresponding to a configuration with two particles at distance $l$ follows
as
\begin{equation}
Y_{n,n+l}=z_{2}^{l}+A_{21}z_{1}^{l}=z_{2}^{l}+z_{2}^{L}z_{1}^{l}=e^{i\gamma
l}+e^{i\gamma\left(  L-l\right)  }. \label{Y_n,n+l}%
\end{equation}

We have to determine that value of $\gamma$ that gives the largest eigenvalue. 
From the fact that the largest eigenvalue $\Lambda=(p\rme^{s}+q\rme^{-s}%
)(e^{i\gamma}+e^{-i\gamma})-2=\mu(s)$ is real, it follows that $\gamma$ is
either real or imaginary. $\gamma$ as function of $s$ is determined by solving
(\ref{A21}) with substitution (\ref{A21per}). For $s$ close to $0$, it turns out that real $\gamma$
corresponds to $s>0$ and imaginary $\gamma$ corresponds to $s<0$. For large
$\rme^{s}\gg1$, $\gamma$ is real and the Bethe amplitude $Y_{n,n+l}$ (\ref{Y_n,n+l}) of the largest
right and left eigenvectors is given by
\begin{equation}
Y_{n,n+l}=2\sin\pi l/L+O(\rme^{-s}) \label{Y12}%
\end{equation}
The origin of the above formula will be explained in the section
\ref{Sec:Nparticles}.

\section{Three particles}
\label{Sec:Threeparticles}

The case of two particles was treated as an introductory pedagogical example.
In this section we study the case of three particles in more detail.

\subsection{Bethe ansatz}

The eigenfunction of the modified rate matrix $\widehat{W}$ for $N=3$
particles on a ring is given by the Bethe ansatz
\begin{equation}
\mbox{$| \, {\varphi}\, \rangle$}=\sum\limits_{1\leq n1<n_{2}<n_{3}\leq L}%
\sum\limits_{\sigma}A_{\sigma}z_{\sigma(1)}^{n_{1}}z_{\sigma(2)}^{n_{2}%
}z_{\sigma(3)}^{n_{3}}|n_{1}n_{2}n_{3}\rangle=\sum\limits_{1\leq
n1<n_{2}<n_{3}\leq L}Y_{n_{1}n_{2}n_{3}}|n_{1}n_{2}n_{3}\rangle
\end{equation}
where $\sigma$ are the $3!=6$ permutations of indices $n_{1},n_{2},n_{3}$.
This eigenfunction has eigenvalues%
\begin{equation}
\Lambda=p\rme^{s}\sum\limits_{i=1}^{3}1/z_{i}+q\rme^{-s}\sum\limits_{i=1}^{3}z_{i}-3
\label{ASEP3EigenValue}%
\end{equation}
and the amplitudes $A_{\sigma}$ are given by requirement that $\varphi$ is an
eigenfunction for particles being nearest neighbours
\cite{Kim95,Schu97,DerridaAppert99}). This yields e.g.
\begin{equation}
\frac{A_{jik}}{A_{ijk}}=-\frac{\alpha_{ji}}{\alpha_{ij}}=-\frac{p\rme^{s}%
+q\rme^{-s}z_{i}z_{j}-z_{j}}{p\rme^{s}+q\rme^{-s}z_{i}z_{j}-z_{i}}
\label{BetheAmplitudesRatio}%
\end{equation}
and similar relations for any other permutation of two indices. From the
periodic boundary conditions it follows that
\begin{equation}
\prod\limits_{i=1,i\neq k}^{3}(-1)\frac{\alpha_{ki}}{\alpha_{ik}}=z_{k}%
^{L}=\prod\limits_{i=1}^{3}\frac{p\rme^{s}+q\rme^{-s}z_{i}z_{k}-z_{k}}%
{p\rme^{s}+q\rme^{-s}z_{i}z_{k}-z_{i}} \label{BetheAnsatz}%
\end{equation}
for all $k$.

As in the previous case $N=2$, from translation invariance
for the ground state components we have $Y_{n_{1}n_{2}%
n_{3}}=Y_{n_{1}+1,n_{2}+1,n_{3}+1}$ we get $z_{1}z_{2}z_{3}=1$, so
we have only two unknown parameters. They must be found from the
Bethe Ansatz equations (\ref{BetheAnsatz}) and the requirement of
maximization of the eigenvalue $\Lambda$. Using
(\ref{BetheAmplitudesRatio}), (\ref{BetheAnsatz}), and
$z_{1}z_{2}z_{3}=1$,  the Bethe amplitudes $Y_{n_{1}n_{2}n_{3}}$
for $n_{1}=n,n_{2}=n+l,n_{3}=n+l+d$ may be
brought to the form
\begin{equation}
Y_{n,n+l,n+l+d}=  A_{123}  [z_{2}^{l}z_{3}^{l+d}+z_{3}^{l}z_{1}^{l+d-L}+z_{1}^{l}%
z_{2}^{l+d}z_{3}^{L}]+A_{213}[1\leftrightarrow2]
\label{ASEP3_RightEigenVector}%
\end{equation}
where $A_{213}=-\alpha_{21}/\alpha_{12}=-(p\rme^{s}+q\rme^{-s}z_{1}z_{2}%
-z_{2})/(p\rme^{s}+q\rme^{-s}z_{1}z_{2}-z_{1}) $, and $[1\leftrightarrow2]$ stands
for the substitution $z_{1}\leftrightarrow z_{2}$ in the expression inside the
first square brackets in (\ref{ASEP3_RightEigenVector}).

\subsection{The limit of large conditioned current}

Consider the limit $\rme^{s}\gg1$, which corresponds to a dynamic restriction on
an atypically large current. Treating $\rme^{-s}$ as a small parameter, we can
write the Bethe equations in the form
\begin{equation}
z_{k}^{L}=1+\rme^{-s}p^{-1}\left(  \sum_{j=1}^{3}z_{j}-3z_{k}\right)  +O(e^{-2s})
\end{equation}
We look for the solution for $z_{k}$ in the form $z_{k}=e^{i\gamma_{k}}$. We
choose $\gamma_{k}$ as
\begin{equation}
\gamma_{k}L=2(k-2)\pi+\alpha_{k}\rme^{-s} +O(e^{-2s}).
\end{equation}
This choice leads to the maximal eigenvalue we are searching (basically, we
have to pick up the set of $z_{k}$ with the largest real part). Note that in
the following we shall also assume the large $L$ limit $L\gg1$. Solving
self-consistently the Bethe equations, we obtain after some algebra
$\alpha_{k}=-3\gamma_{k}/p$, so that
\begin{equation}
\gamma_{k}=\frac{2(k-2)\pi}{L}\left(  1-\frac{3}{pL}\rme^{-s}\right)
+O(e^{-2s}). \label{ASEP3}%
\end{equation}

The respective maximal eigenvalue of the modified rate matrix is given by
\begin{equation}
\mu(s)=p\rme^{s}\sum_{k=1}^{3}z_{k}^{-1}+q\rme^{-s}\sum_{k=1}^{3}z_{k}-3=p\rme^{s}%
J_{0}+O(1)+O(\rme^{-s}) \label{LambdaS}%
\end{equation}
where
\begin{equation}
J_{0}=\sum_{k=1}^{3}\cos\frac{2(k-2)\pi}{L} =  3-\frac{4\pi}{L^{2}} + O(1/L^4).
\label{J0}%
\end{equation}
The respective magnitude of the
enhanced current is $\partial\mu(s)/\partial s=p\rme^{s}J_{0}$.
So indeed the choice $\rme^{s}\gg1$ corresponds to large atypical current.

Substituting the solution of the Bethe equations (\ref{ASEP3})
into (\ref{ASEP3_RightEigenVector}), we obtain the right
eigenfunction $Y_{123}$, which corresponds to a configuration with
$3$ particles, separated by distances $l_{1},l_{2},l_{3}$ where
$\sum l_{j}=L$. The result is (up to an overall
constant)
\begin{equation}
Y_{123}=2\sum_{k=1}^{3}\sin\gamma l_{k}-\rme^{-s}\sin\gamma\sum_{k=1}^{3}%
\cos\gamma l_{k}+O(e^{-2s}), \label{Y123}%
\end{equation}
where $\gamma=\frac{2\pi}{L}\left(  1-3\rme^{-s}/L\right)  $, or, alternatively,
as
\begin{equation}
Y_{123}=8\sin\left(  \frac{\pi}{L}l_{1}\right)  \sin\left(  \frac{\pi}{L}%
l_{2}\right)  \sin\left(  \frac{\pi}{L}l_{3}\right)  +O(\rme^{-s}).
\label{Y123_Product}%
\end{equation}

The left eigenfunction turns out to be the same as the right eigenfunction,
which can be demonstrated either from the Gallavotti-Cohen symmetry or
by the following argument. The transposition of the
matrix $\left(  \widehat{W}(s)\right)  ^{T}$ results in the exchanging
$p\rme^{s}\Longleftrightarrow q\rme^{-s}$ in the original matrix $\widehat{W}(s)$.
Denote the Bethe roots of the transposed matrix as $\tau_{k}$. In can be
verified that the Bethe Ansatz equations and eigenvalue of the transposed
matrix coincide with the old ones (\ref{BetheAmplitudesRatio}),
(\ref{ASEP3EigenValue}), after substitution $\tau_{k}\rightarrow1/z_{k}$ for
all $k$. Under these substitutions, the amplitudes of the left eigenfunction
become $Y_{123}^{\LEFT}=\sum\limits_{\sigma}A_{\sigma}z_{\sigma(1)}^{-n_{1}%
}z_{\sigma(2)}^{-n_{2}}z_{\sigma(3)}^{-n_{3}}$. However, note from
(\ref{ASEP3}) that $z_{1}=z_{3}^{-1}$ ,and $z_{2}=1,$ up to corrections
$O((\rme^{-s})^{2})$. Therefore, by relabelling the roots $z_{1}\leftrightarrow
z_{3}$ we obtain $Y_{123}^{\LEFT}=\sum\limits_{\sigma}A_{\sigma}z_{\sigma
(1)}^{n_{1}}z_{\sigma(2)}^{n_{2}}z_{\sigma(3)}^{n_{3}}=Y_{123}^{\RIGHT}$.

With that in mind, we are in a position to construct the \textit{effective
stochastic dynamics}, defined by (\ref{EffectiveDynamicsRate}) conditioned on
large flux.

To this end, notice first that the stationary probabilities to find the
configuration of particles at distances $l_{1},l_{2},l_{3}$, is given by
(\ref{EffectiveDynamicsStationaryState})
\begin{equation}
P^{\eff}(l_{1},l_{2},l_{3})=\left(  Y_{123}\right)  ^{2}/Z
\end{equation}
where $Z=\left(  \sum_{l_{1},l_{2},l_{3}}\left(  Y_{123}\right)  ^{2}\right)
=3L^{2}$ is the normalization factor. With (\ref{Y123_Product}) it is readily
checked that

\begin{itemize}
\item \textbf{(A)} the configuration with $l_{1}=l_{2}=l_{3}=L/3$ is the most
probable one

\item \textbf{(B)} all configurations with \textit{any} single finite distance
$l_{k}$. i.e., where $l_{k}/L\ll1$, and arbitrary other distances $l_{j}$ have
vanishing probability
\end{itemize}

Indeed, this is consistent with the intuition that the configurations where
two particles are coming near each other (close on lattice scale) should be
avoided in the effective stochastic dynamics, because it would tend to cause a
current reduction due to hard-core repulsion.

The rates for the \textit{effective stochastic dynamics} are given by
(\ref{EffectiveDynamicsRate}). Namely, the rate for hopping to the right of
the second particle at position $x_{2}$ in the configuration with consecutive
distances $l_{i}=x_{i+1}-x_{i},$ is given by
\begin{equation}
W_{l_{1}+1,l_{2}-1,l_{3}|l_{1}l_{2}l_{3}}^{\eff}=p\rme^{s}\left[  \frac
{\sin\left(  \frac{\pi}{L}\left(  l_{1}+1\right)  \right)  \sin\left(
\frac{\pi}{L}\left(  l_{2}-1\right)  \right)  }{\sin\left(  \frac{\pi}{L}%
l_{1}\right)  \sin\left(  \frac{\pi}{L}l_{2}\right)  }+O(\rme^{-s})\right]  .
\end{equation}
(remember that $W_{C^{\prime}C}$ is rate of change $C\rightarrow C^{\prime}$).
At short distances $l_{1}/L\ll1$ or $l_{2}/L\ll1$ the major contribution to
this expression in square brackets is calculated by an expansion
$\sin\varepsilon\approx\varepsilon$. We obtain that the rates of hopping for a
two particles at small distances $d$ , $d/L\ll1$ are given by
\begin{equation}
W_{..d..,..d+1..}^{\eff}\approx p\rme^{s}\left[  \frac{d}{d+1}\right]  ,\text{
}d=1,2,... \label{UnfavourableRates}%
\end{equation}%
\begin{equation}
W_{..d+1..,..d..}^{\eff}\approx p\rme^{s}\left[  \frac{d+1}{d}\right]  ,\text{
}d=1,2,... \label{FavourableRates}%
\end{equation}
At the same time, the rates of hopping to the left in the effective dynamics
will have the prefactor $q\rme^{-s}$, and in the limit $\rme^{-s}\ll1$ can be
neglected. We see that the rates (\ref{FavourableRates}) driving two particles
apart are larger than those driving them together (\ref{UnfavourableRates}),
and therefore the configurations with two particles forming a cluster are
difficult to reach. However, in any finite system $L<\infty$, even in the
limit $\rme^{-s}\rightarrow0$, configurations with particle distances ($l_{i}=1$)
are possible.

It is instructive at this point to compare the probability of forming of a
cluster in the model with the effective dynamics and in the original ASEP. In
the model with the effective dynamics the single-cluster probability is
\begin{equation}
P_{3}^{\eff}=3Y_{l_{1}=1,l_{2}=1,l_{3}=L-2}^{2}/(3L^{2})\rightarrow\frac
{2^{8}\pi^{6}}{L^{8}}\text{ for }L\gg1,
\end{equation}
and the probability of a cluster of two particles is%
\begin{equation}
P_{2}^{\eff}=\frac{3}{3L^{2}}\sum_{l_{2}=2}^{L-3}Y_{l_{1}=1,l_{2}%
,l_{3}=L-1-l_{2}}^{2}\rightarrow2^{6}\frac{\pi^{2}}{L^{2}}L\langle\sin
^{4}x\rangle/(L^{2})=\frac{24\pi^{2}}{L^{3}}\text{ for }L\gg1.
\end{equation}
On the other hand in the original unconstrained ASEP with $3$ particles on a
ring, the probability to form a single cluster is $P_{3}^{ASEP}=L/\binom{L}%
{3}\rightarrow6/L^{2}$ for $L\gg1$, and the probability to form a cluster of
two particles is $P_{2}^{ASEP}=L(L-4)/\binom{L}{3}\rightarrow6/L$ for $L\gg1$.
\ So we see that the clustering is strongly suppressed in the effective
dynamics, with respect to the unforced ASEP, which is natural to expect in \ a
system conditioned to produce a large current. On the other hand, the
stationary probability of equidistant state $l_{1}=l_{2}=l_{3}=L/3$ (the most
probable state in the effective dynamics)
\begin{equation}
P_{\text{equid.}}^{\eff}=Y_{l_{1}=l_{2}=l_{3}=L/3}^{2}/(3L^{2})=9/L^{2}
\label{Peff_equid}%
\end{equation}
is only $4.5$ times larger than in the unconstrained ASEP $P_{equid}%
^{ASEP}=\frac{L}{3}/\binom{L}{3}\rightarrow2/L^{2}$ for $L\gg1$.

\begin{figure}[ptb]
\unitlength=1mm \makebox(60,60)[cc]
{\psfig{file=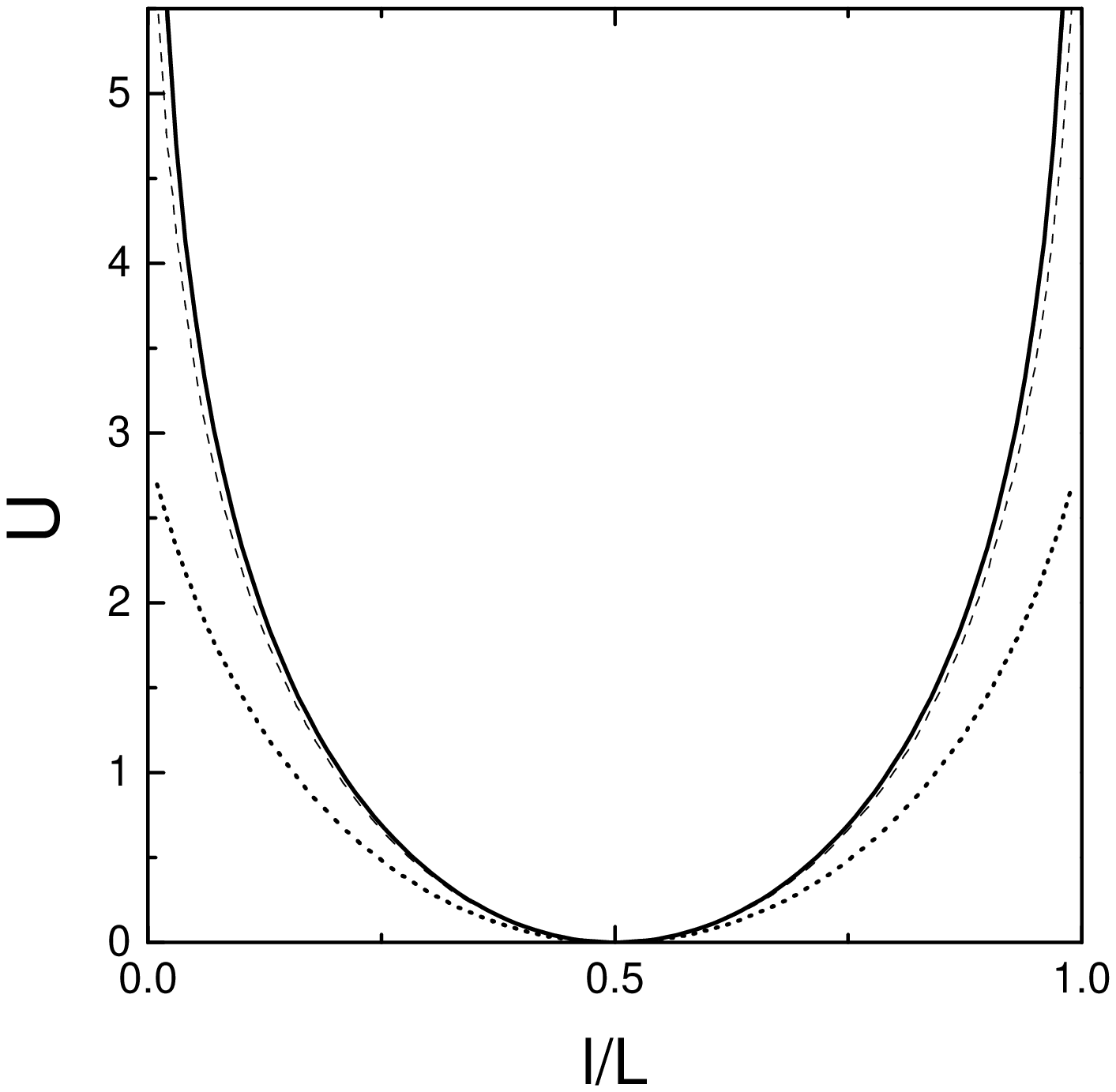,width=60mm}}
\caption{Effective Potential $U$ felt by
two particles on a ring of $L=100$ sites. The short-distance configurations
($l/L=0,1$) are unlikely. The solid, dashed and dotted lines correspond to
$\rme^{-s}=0,0.5,0.9$ respectively. }%
\label{Fig_Ufor2}%
\end{figure}

Finally, we can define an \textit{effective potential }$\ U$, felt by the
particles evolving according to the effective stochastic dynamics, in obvious
analogy with equilibrium thermodynamics, by%
\begin{equation}
\beta U(C)=-\ln P^{\eff}(C)=-2\ln\langle\mu_{1}(s)|C\rangle,
\label{EffectivePotential}%
\end{equation}
where $\beta$ is the effective inverse temperature. Consider first the case of
two particles, where the amplitude $\langle\mu_{1}(s)|C\rangle$ for the
configuration of two particles at distance $l$ is given by $Y_{12}=\sin\pi
l/L+O(\rme^{-s})$. Hence
\begin{equation}
\beta U(l)=-2\ln Y_{12}=-2\ln\sin\frac{\pi l}{L}+const
\end{equation}
shown on Fig.\ref{Fig_Ufor2} (solid line). By numerical evaluation of the
Bethe ansatz equation we have also determined the effective potential for
finite $s$. Analogously the effective potential for three particles takes
form
\begin{equation}
\beta U=-2\ln Y_{123}%
\end{equation}
where $Y_{123}$ is given by formula (\ref{Y123}). Note that below we shall use
the freedom in defining the potential by eliminating a constant, connected to
the normalization of the $P^{\eff}(C)$.

\subsection{First order corrections in $\rme^{-s}$ to the effective dynamics}

Using (\ref{Y123}), one can obtain first order correction to the effective
dynamics. E.g., for the hopping rates to the right (\ref{UnfavourableRates})
we get, assuming $d/L\ll1$%
\begin{equation}
W_{\ldots d \ldots,\ldots d+1 \ldots}^{\eff}(s)= p\rme^{s}\frac{d}{d+1}
\left(  1-\frac{\rme^{-s}}{d(d+1)}\frac{1+5\cos2\pi u}{1-\cos2\pi u}
+ O(\rme^{-2s})   \right)  ,\text{ }
\label{UnfavourableRates_s}%
\end{equation}
where $u=l_{1}/L$, for a configuration with two particles at small distance
$d$.

For generic configurations with particles separated by distances which scale
with $L$, we get, by expanding the generic expression
(\ref{EffectiveDynamicsRate}) with substitution (\ref{Y123}) in Taylor series
with respect to $1/L$ and $\rme^{-s}$ and keeping the first order terms
\begin{equation}
W_{l_{1}+1,l_{2}-1,l_{3}|l_{1}l_{2}l_{3}}^{\eff}(s)=p\rme^{s}\left(  1+\frac{2\pi
}{L}\frac{\left(  \cos\frac{2\pi l_{1}}{L}-\cos\frac{2\pi l_{2}}{L}\right)
}{\sum_{k=1}^{3}\sin\frac{2\pi l_{k}}{L}}+\rme^{-s}\frac{2\pi}{L^{2}}\frac
{\Phi\left(  \frac{l_{2}}{L}\right)  -\Phi\left(  \frac{l_{1}}{L}\right)
}{\sum_{k=1}^{3}\sin\frac{2\pi l_{k}}{L}}\right)  , \label{BulkRates_s}%
\end{equation}
where $\Phi\left(  u\right)  =\frac{\partial}{\partial u}\left[
(1+3u)\sin2\pi u\right]  $. The last term in the brackets gives the
contribution due to finite $s$. The relative effective potential differences
is given the ratio between the middle and the last term in the brackets,
\begin{equation}
\frac{\Delta U(s)-\Delta U(s\rightarrow\infty)}{\Delta U(s\rightarrow\infty
)}=\frac{\rme^{-s}}{L}\frac{\Phi\left(  \frac{l_{2}}{L}\right)  -\Phi\left(
\frac{l_{1}}{L}\right)  }{\cos\frac{2\pi l_{1}}{L}-\cos\frac{2\pi l_{2}}{L}}.
\label{PotentialDifferences_s}%
\end{equation}
As the system size $L$ becomes infinite, the first order corrections to the
hopping rates due to finite $s$ survive for configurations with finite (on
lattice scale) distances as (\ref{UnfavourableRates_s}) shows, because it does
not have $1/L$ dependence in it. On the contrary, for configurations with all
inter-particle distances scaling with system size $0<l_{i}/L<1$, for
$L\rightarrow\infty$, the corrections due to finite $s$ disappear, see
(\ref{BulkRates_s}), (\ref{PotentialDifferences_s}). Another example of this can
be seen in Fig.\ref{Fig_Ufor2}.

\section{Arbitrary number of particles in the large current limit}
\label{Sec:Nparticles}

\subsection{Largest eigenvalue and associated eigenvector}

The arguments, discussed in previous sections, are readily generalized to the
case of an arbitrary number of particles $N$. We shall treat the large current
limit $\rme^{s}\rightarrow\infty$ only. Note that unless stated otherwise, the
system size $L$ is arbitrary. Note also that in the limit $\rme^{s}%
\rightarrow\infty$ , to avoid the divergences, the modified rate matrix
(\ref{ModifiedRateMatrix}) has to be rescaled by the factor $\rme^{-s}$ as
$\rme^{-s}\widehat{W}(s)$.

For finite $s$ the eigenfunction of the modified rate matrix is given by the Bethe Ansatz
\cite{Schu97,DerridaAppert99},
\begin{equation}
\mbox{$| \, {\varphi}\, \rangle$}=\sum\limits_{1\leq n1<n_{2}<n_{3}\leq L}%
\sum\limits_{\sigma\in S_{N}}A_{\sigma}z_{\sigma(1)}^{n_{1}}z_{\sigma
(2)}^{n_{2}}...z_{\sigma(N)}^{n_{N}}|n_{1}n_{2}...n_{N}\rangle=\sum
\limits_{1\leq n1<n_{2}<n_{3}\leq L}Y_{12..N}|n_{1}n_{2}...n_{N}\rangle
\end{equation}
where the coefficients $A_{\sigma}$ are given by
\begin{equation}
\frac{A_{...\sigma(ij)...}}{A_{...ij....}}=-\frac{\alpha_{ji}}{\alpha_{ij}%
}=-\frac{p\rme^{s}+q\rme^{-s}z_{i}z_{j}-z_{j}}{p\rme^{s}+q\rme^{-s}z_{i}z_{j}-z_{i}}%
\end{equation}
and the Bethe roots $z_{j}$ satisfy
\begin{equation}
\prod\limits_{i=1,i\neq k}^{N}(-1)\frac{\alpha_{ki}}{\alpha_{ik}}=z_{k}%
^{L}=(-1)^{N-1}\prod\limits_{i=1}^{N}\frac{p\rme^{s}+q\rme^{-s}z_{i}z_{k}-z_{k}%
}{p\rme^{s}+q\rme^{-s}z_{i}z_{k}-z_{i}} \label{BetheAnsatzN}%
\end{equation}
for arbitrary $N$.

In the limit $\rme^{s}\rightarrow\infty$ the solution of the Bethe equations
corresponding to the largest eigenvalue of
the modified rate matrix  has the form
$z_{k}=\exp(i\gamma_{k})$, where
\begin{equation}
\gamma_{k}=\frac{2(k-\frac{N+1}{2})\pi}{L}\text{, \ }k=1,2,...N
\label{gamma_k}%
\end{equation}
generalizing (\ref{ASEP3}). The maximal eigenvalue of the rescaled modified
rate matrix $\rme^{-s}\widehat{W}(s)$ and the respective rescaled current are given
by
\begin{equation}
\lim_{s\rightarrow\infty}\rme^{-s}\mu(s)=p\sum_{k=1}^{N}z_{k}^{-1}=J_{0}
\label{LambdaSN}%
\end{equation}
where
\begin{equation}
J_{0}=p\sum_{k=1}^{N}\cos\frac{2(k-\frac{N+1}{2})\pi}{L} = p\frac{\sin\pi\rho}{\sin\frac{\pi}{L}}. \label{J0N}%
\end{equation}
The coefficients
$A_{\sigma}$ are simplified to
\begin{equation}
A_{\sigma}=sgn(\sigma)
\end{equation}
where $sgn(\sigma)$ denotes a signature of permutation $\sigma$. The
limiting expression for $Y_{12..N}$ can then be rewritten in the equivalent form
\begin{equation}
\left.  Y_{12..N}\right\vert _{e^{s}\rightarrow\infty}=\sum\limits_{\sigma\in
S_{N}}sgn(\sigma)z_{1}^{\sigma(n_{1})}z_{2}^{\sigma(n_{2})}...z_{N}%
^{\sigma(n_{N})}.
\end{equation}

In the following we drop the subscript indicating that we are taking the limit of large $s$.
Using the Leibniz formula for the determinant, we can rewrite the above
expression as a Slater determinant
\begin{equation}
Y_{12..N} =\det
\begin{vmatrix}
z_{1}^{n_{1}} & z_{1}^{n_{2}} & ... & z_{1}^{n_{N}}\\
z_{2}^{n_{1}} & z_{2}^{n_{2}} & ... & z_{2}^{n_{N}}\\
... & ... & ... & ...\\
z_{N}^{n_{1}} & z_{N}^{n_{2}} & ... & z_{N}^{n_{N}}%
\end{vmatrix}
\label{SlaterDeterminant}%
\end{equation}
For the choice of Bethe roots (\ref{gamma_k}) this determinant  takes the form
\begin{equation}
Y_{12..N} =\exp\left[
-(N-1)\sum_{j=1}^{N}i\pi x_{j}\right]  \prod\limits_{\substack{l,k\\1\leq
l<k\leq N}}\left(  \rme^{2i\pi x_{k}}-\rme^{2i\pi x_{l}}\right)
\end{equation}
where $x_{k}=n_{k}/L$, which becomes simply
\begin{equation}
Y_{12...N}=2^{\binom{N}{2}}\prod\limits_{\substack{l,k\\1\leq l<k\leq N}%
}\sin\pi\frac{n_{k}-n_{l}}{L}. \label{SinVandermonde}%
\end{equation}
The amplitudes $Y_{12...N}$ at this stage do not have a probabilistic meaning
and are not normalised.

For $N=2,3$ we recover the expressions
(\ref{Y12}), (\ref{Y123_Product}) up to  finite $s$
corrections. By extending the arguments given after (\ref{Y123_Product}) we readily
verify that the left maximal eigenvector is given by the same expression
(\ref{SinVandermonde}). Consequently, the stationary probabilities of particle
configurations in the effective dynamics are equal to the normalized square of the
amplitudes (\ref{SinVandermonde}),
\begin{equation}
P_{12...N}^{\eff}=\frac{1}{K_{N,L}}\prod\limits_{\substack{l,k\\1\leq l <k\leq
N}}\sin^{2}\pi\frac{n_{k}-n_{l}}{L},
\label{StationaryProbabilitiesSinVandermonde}%
\end{equation}
where
\begin{equation}
K_{N,L}=2^{-N(N-1)}L^{N} \label{NormalizationK_N,L}%
\end{equation}
is the free fermion normalization factor \cite{Gaud81}. Notice that in this result,
which is exact for any system size $L\geq 1$ the
particle number $N$ is subject only to the exclusion rule $0 \leq N \leq L$. 
As detailed in section \ref{sec:cue} further below, this formula is 
similar to the distribution of eigenvalues in the circular unitary 
ensemble of random matrices.

Let us analyze our main results (\ref{SinVandermonde}),
(\ref{StationaryProbabilitiesSinVandermonde}). The stationary probabilities
depend on the scaled macroscopic distances $x_k$. For $N$ fixed and finite
they satisfy the following properties in the thermodynamic limit $L\to \infty$:%
\begin{equation}
\text{\textbf{(A) \ }}\max Y_{12...N}\text{ is reached for }|x_{k+1}%
-x_{k}|=\frac{1}{N}\text{ for all }k \label{propertyA}%
\end{equation}%
\begin{equation}
\text{\textbf{(B) \ }}Y_{12...N}=0\text{ for }\prod\limits_{k=1}^{N}\left(
x_{k+1}-x_{k}\right)  =0 \label{propertyB}%
\end{equation}
The property \textbf{(A)} indicates that the most probable
configuration of the system is an equidistant configuration.
The property \textbf{(B)} is evident from (\ref{SinVandermonde}) and implies
that particles cannot be found with finite probability at distances which are finite
on lattice scale (or in fact in any non-macroscopic distance $o(L)$).

\subsection{Effective stochastic process}

Now we describe the effective stochastic process corresponding to the limit
$\rme^{s}\rightarrow\infty$. The effective hopping rates are given by
(\ref{EffectiveDynamicsRate}): The move of the $k$-th particle, located at
position $n_{k}=Lx_{k}$, to consecutive position $n_{k}+1$, provided it is
vacant, has the rate%
\begin{equation}
\lim_{s\rightarrow\infty}\rme^{-s}p^{-1}W_{C^{\prime}C}^{\eff}=
\frac{Y_{12..N}^{\prime}}{Y_{12..N}}=\prod\limits_{l\neq k}
\frac{\sin\pi\frac{n_{k}-n_{l}+1}{L}}
{\sin\pi\frac{n_{k}-n_{l}}{L}}.
\label{e^-sW}%
\end{equation}
where the product is over $l$ and $k$ as above.
The stationary state probabilities
with respect to the above rates are given by
(\ref{StationaryProbabilitiesSinVandermonde}).

The effective potential of interaction, given by (\ref{EffectivePotential}),
has the form of a long-range potential
\begin{equation}
U(x_{1},x_{2},...,x_{N})=-2\sum_{\substack{l,k\\l\neq k}}\log|\sin\pi
(x_{l}-x_{k})|. \label{U}%
\end{equation}
Note that the summation in (\ref{U}) extends over all $i\neq j$. Note also
that due to the ring geometry and the fact that $|\sin(\pi\pm\pi c)|=|\sin\pi
c|$, the differences $x_{l}-x_{k}$ in (\ref{U}) can be
substituted by the smallest distances between particles $l,k$ on a ring. 
As for the stationary distribution of the particles 
\eqref{StationaryProbabilitiesSinVandermonde}, this effective potential $U$ is 
similar to the potential that appears for eigenvalues in random unitary matrices 
that perform Dyson's Brownian motion over the unitary group 
$U(N)$ (see details below in section \ref{sec:cue}).

For any finite number of particles $N/L \ll 1$, the long range effective
potential $U$ attains in the thermodynamic limit $L\to\infty$
singularities for any configuration with finite distances
(on lattice scale) between the neighboring particles.
For the corresponding hopping rates of the $k$-th particle
we find to leading order in $1/L$
\begin{align}
\lim_{s\rightarrow\infty}\rme^{-s} p^{-1} W_{C^{\prime}C}^{\eff}  &  =
1+\frac{1}%
{L}\frac{\partial}{\partial x_{k}}\ln Y_{12..N}=\nonumber\\
&  =1+\frac{\pi}{L}\sum_{l\neq k}\cot\pi(x_{k}-x_{l}).
\label{GeneralEffectiveHopRates}%
\end{align}
This is obtained by expanding (\ref{e^-sW}) in a Taylor series
with respect to $1/L$. Such dynamics correspond to an enhancement
of the hopping bias whixh vanishes at macroscopic distances, but which
is strong at microscopic distances.

It is instructive to compare some configurational probabilities in the ASEP
with the effective model for finite \textit{fraction} of particles. For the
half-filled lattice $N/L=1/2$, the probability of the equidistant particle
state $0,1,0,1,...$ in the ASEP (without conditioning) is given by
$P_{ASEP}^{equidist}(..0101..)=2/\binom{L}{L/2} \sim 2^{-L}$ for large
$L$. In the effective dynamics, the probability of the same state\ (the state
of largest probability ) is expressed, after some algebra, in terms of the
normalization factors (\ref{NormalizationK_N,L}) as $P^{\eff}%
(..0101..)=2K_{L/2,L/2}/K_{L/2,L}=2^{1-\frac{L}{2}}$. So we see that
\begin{equation}
P^{\eff}(..0101..)=\sqrt{P_{ASEP}^{equidist}(..0101..)},
\end{equation}
up to terms which grow in size slower than exponentially.

This calculation can
be easily extended for arbitrary finite filling fraction of the form $N/L=\nu
_{F}=1/2,1/3,..$. The ASEP weight of the equidistant configuration for $L\gg1$
is $P_{ASEP}^{equidist}(100...\underbrace{100...1}_{1/\nu_{F}}00..)=\nu
_{F}^{-1}\binom{L}{L\nu_{F}}^{-1}\approx \rme^{L\left(  v_{F}\ln\nu_{F}%
+(1-v_{F})\ln(1-\nu_{F})\right)  }$, and the respective weight under the effective
dynamics is given by $P^{\eff}(\nu_{F})=K_{L\nu_{F},L\nu_{F}}/(\nu_{F}%
K_{L\nu_{F},L})=\nu_{F}^{L\nu_{F}-1}\approx \rme^{Lv_{F}\ln\nu_{F}}$. Comparing
the two probabilities, we obtain, assuming $L\gg1$:
\begin{equation}
\frac{P_{ASEP}^{equidist}(100...\underbrace{100...1}_{1/\nu_{F}}00..)}%
{P^{\eff}(\nu_{F})}=e^{L(1-v_{F})\ln(1-\nu_{F})}.
\end{equation}
Moreover, using the connection to the quantum free fermion model at zero
temperature (see Conclusions), various correlation functions for the effective
stochastic dynamics can be obtained from the well-known free fermion quantum
expectation values. The particle-particle correlation function in
the effective dynamics $\langle\hat{n}_{k}\hat{n}_{k+m}\rangle$ is equal to
the ground state expectation value $\langle(1-\sigma_{k}^{Z})(1-\sigma
_{k+m}^{Z})\rangle/4$ in the free fermion problem and is given in the
thermodynamic limit $L\rightarrow\infty,N/L\rightarrow\rho$ by
\begin{equation}
\langle\hat{n}_{k}\hat{n}_{k+m}\rangle^{\eff}=\rho^{2}-\frac{1}{\pi^{2}}%
\frac{\sin^{2}m\pi\rho}{m^{2}} \label{particle-particleCorrelation}%
\end{equation}
see e.g.\cite{KorepinTMP93}. Note that in the unconstrained ASEP there are no
correlations in the stationary state $\langle\hat{n}_{k}\hat{n}_{k+m}%
\rangle_{ASEP}=\langle\hat{n}_{k}\rangle\langle\hat{n}_{k+m}\rangle=\rho^{2}$
for any $m>0$. For half-filling $\rho=1/2$\ and nearest-neighbour
correlations we obtain $\langle11\rangle^{\eff}=$ $\langle00\rangle
^{\eff}=1/4-1/\pi^{2}\approx0.149$, and $\langle10\rangle^{\eff}=$
$\langle01\rangle^{\eff}=1/4+1/\pi^{2}\approx0.351$ for the effective dynamics
while $\langle11\rangle=\langle00\rangle=\langle10\rangle=\langle
01\rangle=1/4$ for the original ASEP.  This again shows that the effective
dynamics of ASEP conditioned on high current avoids building clusters, as is
expected. Interestingly, for $\rho=1/2$ the effective particle particle
correlations (\ref{particle-particleCorrelation}) at even distances $m=2n$
coincide with those for ASEP, $\langle\hat{n}_{k}\hat{n}_{k+2n}\rangle
^{\eff}=\langle\hat{n}_{k}\hat{n}_{k+2n}\rangle^{ASEP}=1/4$ while the
odd-distance correlations decay algebraically $\langle\hat{n}_{k}\hat
{n}_{k+(2n+1)}\rangle^{\eff}=1/4-1/(\pi^{2}(2n+1)^{2})$.

\subsection{Circular unitary ensemble and Dyson's Brownian motion}
\label{sec:cue}

Exclusion processes on $\mathbb{Z}$ are already known to share 
common features \cite{Joha00,borodin,peche} with random matrices, 
such as the determinantal structure, the Airy fluctuations and the 
Tracy Widom distribution for different quantities in the TASEP. Howver, much less 
is known about the connection between the ASEP on a ring with periodic 
boundary conditions and random matrix ensembles. 

The circular unitary ensemble is defined as the Haar measure on the 
unitary group $\mathcal{U}(N)$. This group is defined as the set of 
complex matrices $U$ such that $U^\dagger U=I$. This Lie group can 
be equipped with a \emph{uniform} measure, called the \emph{Haar measure}. 
Almost every matrix $U\in \mathcal{U}(N)$ can be diagonalized and the Haar 
measure induces a new measure on the eigenvalues, which contains a 
determinantal-like interaction between them. All eigenvalues of a unitary 
matrix lie on the unit circle $\mathbb{U}$ and can be written as 
$\rme^{i\theta_k}$ where $0\leq \theta_k < 2\pi$. The density probability of 
eigenvalues is 
\begin{equation}
\label{haar}
P_\text{CUE}( \theta_1,\ldots,\theta_N) = 
\frac{1}{Z_N} \prod_{1\leq k < l \leq N} \left| \rme^{i\theta_k}-\rme^{i\theta_j} \right|^2 
= \frac{1}{Z'_N} \prod_{1\leq k < l \leq N} \sin^2\left(\frac{\theta_k-\theta_l}{2} \right),
\end{equation}
which is precisely the formula \eqref{StationaryProbabilitiesSinVandermonde}, 
up to the fact that the $n_j/L$ are rational numbers due to the discrete structure 
of the underlying lattice.

This analogy can be developed further if one considers Dyson's 
Brownian motion \cite{dyson} on the unitary group $\mathcal{U}(N)$. 
The infinitesimal evolution of the process $U_t$ on $\mathcal{U}(N)$ starting 
at the identity matrix is given by (It\^{o} form):
\begin{equation}
dU_t = i dH_t U_t - \frac{1}{2} U_t dt
\end{equation}
and can also be seen as the exponential
\begin{equation}
U_t = \rme^{iH_t}
\end{equation}
where $H_t$ is a random hermitian matrix whose independent entries 
perform independent Brownian motions. The independent diffusions of 
the coefficients induce interacting diffusions on the eigenvalues 
$\rme^{i\theta_k(t)}$ of the matrix $U_t$:
\begin{equation}
\label{eq:dysonev}
d\theta_k = \sum_{l\neq k} \cot\left( \frac{\theta_l-\theta_k}{2}\right) dt+ dB^{(k)}_t
\end{equation}
where the $B_t^{(k)}$ are $N$ independent Brownian motions. Up to 
a global common drift term, the interaction term is similar to the one obtained in 
\eqref{GeneralEffectiveHopRates} for the ASEP conditioned to produce a 
large current and it derives from the same long-range potential $U$ as 
defined in \eqref{U}. The invariant measure of the Dyson Brownian 
motion over $\mathcal{U}(N)$ is precisely the Haar measure \eqref{haar}.

Although the similarity is striking, a deep understanding of the connection 
between both models and the existence of further similarities (dynamical 
fluctuations, corrections in $\rme^{-s}$) are still missing.

\subsection{Vandermonde determinant limit and clustering probabilities}

Let us consider a special case of finite number of particles $N$, which form a
\emph{\ single cluster}, where all distances between particles are small, i.e.
$|x_{k}-x_{l}|\ll1$ for all $k,l$. In this case the amplitude of the wave
function (\ref{SinVandermonde}) becomes proportional to the Vandermonde
determinant
\begin{equation}
Y_{12...N} \propto \prod
\limits_{1\leq l<k\leq N}\left(  x_{k}-x_{l}\right)  =\det%
\begin{vmatrix}
1 & x_{1} & ... & x_{1}^{N-1}\\
1 & x_{2} & ... & x_{2}^{N-1}\\
.. & .. & ... & ...\\
1 & x_{N} & ... & x_{N}^{N-1}%
\end{vmatrix}
\end{equation}
The respective stationary probabilities of the configurations are proportional
the square of the Vandermonde determinant,
\begin{equation}
 P_{x_{1}x_{2}...x_{N}} \propto \prod_{1\leq l<k\leq
N}\left(  x_{k}-x_{l}\right)^{2}%
\end{equation}
For instance, the ratio of probabilities of two similar single cluster
configurations $C,C^{\prime}$ where $C^{\prime}$ differs from $C$ by rescaling
all particle-particle distances by a common factor $r,$ namely, $|x_{k}%
^{\prime}-x_{l}^{\prime}|=r|x_{k}-x_{l}|$ for all $k,l$, is given by
\begin{equation}
\frac{P_{x_{1}^{\prime}x_{2}^{\prime}...x_{N}^{\prime}}^{\eff}}{P_{x_{1}x_{2}...x_{N}}^{\eff}}
=r^{N(N-1)},
\end{equation}
which grows much faster than exponentially with number of particles $N$ (an
exponential growth would be expected for uncorrelated particles, restricted
to smaller volume). E.g. single cluster configurations of $3$ particles ($10$
particles) are $2^{6}$ times ($2^{90}$ times) less likely than respective
similar configurations of double size $r=2$.

\subsection{Gap in the spectrum of the effective Hamiltonian}

The expression (\ref{SinVandermonde}) gives the components of the maximal
eigenvector of the effective Markov rate matrix $W^{\eff}$ which corresponds to
the stationary state. Let us find the next largest eigenvalue of $W^{\eff}$
which determines the longest relaxation time. Following our introductory
exposition (see also \cite{Simo09}) all eigenvalues of  $W^{\eff}$ are given
by $\Lambda_{j}(s)-\mu(s)$
where $\Lambda_{j}(s)$ are eigenvalues of the modified rate matrix $\widehat{W}(s)$
(\ref{ModifiedRateMatrix}). In the present case the next largest eigenvalue of $\widehat{W}(s)$ is
given by the set of Bethe roots for the largest one (\ref{gamma_k}), where the
last root $z_{N}=\exp\left(  i\frac{N-1}{L}\right)  $ is replaced by the root
$z_{N}^{\prime}=\exp\left(  i\frac{N+1}{L}\right)  $. Such a choice
corresponds to a minimal modification in the set (\ref{gamma_k}) and it is
inherited from the underlying free fermion quantum system.

Computing the difference between the eigenvalues $\Lambda=\mu(s)$ and
$\Lambda^{\prime}$ given by the set (\ref{gamma_k}) and the new set, we obtain
to leading order in $\rme^{s}$ the real part of the spectral gap of $W^{\eff}$, as
\begin{align}
\operatorname{Re}\left[  \mu(s)-\Lambda^{\prime}\right]   &  \sim p\rme^{s}%
(\cos\pi(\rho-\frac{1}{L})-\cos\pi(\rho+\frac{1}{L}))
\nonumber\\
&  =2 p \rme^{s}\sin{\frac{\pi}{L}}\sin\pi\rho \label{SpectrumGap}%
\end{align}
where $\rho=N/L$ is the particle density. Since our choice of Bethe roots
corresponds to $\rme^{s}\rightarrow\infty$ limit, we see that for large $L$ the gap in the
rescaled effective Hamiltonian is inversely proportional to the system size
$\rme^{-s}W^{\eff} = O(1/L)$. Note that in the unconstrained ASEP the real part
of the spectrum gap scales as $O(1/L^{3/2})$. We conclude that the long-range
interaction of the effective process
leads to a marked reduction of the
longest relaxation time (which is the inverse spectral gap). 

\section{Conclusions}
\label{Conclusions}

We have shown that conditioning the ASEP on a long period of strongly enhanced
current can be described in terms of an effective exclusion process with long range
interactions. Numerical evaluation of Bethe ansatz equations and
perturbation theory in $\mbox{\rm e}^{-s}$ show that the explicit asymptotic results
that we have obtained for very large fluxes are rather robust with
respect to the value of $s$, at least for finite number of particles $N/L\ll1$
and large system sizes $L$. Indeed, as can be seen in Fig.\ref{Fig_Ufor2}, the
effective two-particles potential $U$ for $\rme^{-s}\leq0.5$ is difficult to
distinguish from the limit $\rme^{-s}=0$ for $L=100$. For $L=1000$, the
differences from the limiting case would become visible in Fig.\ref{Fig_Ufor2}
only at values $\rme^{-s}\gtrsim0.95$ and higher (curves not shown). The
case $\rme^{-s}=1$ ($s=0$) corresponds to the usual ASEP dynamics on a ring
without conditioning and presents a singular limit.

The hopping rates corresponding to the limiting effective dynamics have been
computed explicitly, see (\ref{e^-sW}). They can be interpreted as arising from an
equilibrium potential that depends only on the interparticle distance. We have
computed various space-time properties of the dynamics conditioned to carry a
large current for a long time, i.e., of the quasi-stationary distribution. Density
correlations decay with a power law and have a periodic modulation
(\ref{particle-particleCorrelation}).
Temporal correlations decay with a dynamical exponent $z=1$ rather than with
the KPZ exponent $z=3/2$ that characterizes the typical behaviour of the ASEP.
Hence in the extreme-current regime the process is in a different universality
class, indicating a phase transition as one enters into the regime of strongly
non-typical behaviour. Non-analytic behaviour of physical observables in the
regime of strong non-typical behaviour has been shown to occur also for
atypical small currents and in the zero-range process for high currents.

Our approach highlights a surprising analogy between the \textit{classical}
stochastic problem we have started from with an intrinsically \textit{quantum}
problem of $N$ free fermions on a periodic lattice. This analogy goes far 
beyond the quantum formalism used in section \ref{Approach} to describe the 
stochastic classical dynamics. In fact, the expression
(\ref{SlaterDeterminant}), and, correspondingly, also (\ref{SinVandermonde})
is a Slater determinant describing the wave function of a system of free
fermions, the so-called $XX0$ model,%
\begin{equation}
\label{HXX0}H_{XX0}=-\sum_{n=1}^{L}\left(  \sigma_{n}^{+}\sigma_{n+1}%
^{-}+\sigma_{n+1}^{+}\sigma_{n}^{-}\right)
\end{equation}
with periodic boundary conditions $L+1\equiv1$. The probabilities of the
\textit{classical} particle configurations under effective dynamics given by
$P^{\eff}(C)=\langle\mu_{1}(s)|C\rangle\langle C|\mu_{1}(s)\rangle$, coincide,
up to terms of order $\rme^{-s}$, with \textit{quantum} expectation values
for the free fermion problem because of the presence of both left 
and right eigenvectors, whereas the unconditioned probability involves 
only the right eigenvector. The reason why the free fermion Hamiltonian
has a common eigenvector with the modified rate matrix in the
limit $\mbox{\rm e}^{-s}\to0$ is simple: In this limit, the modified rate
matrix becomes $\rme^{-s}\widehat{W}(s)=\sum_{n=1}^{L}\left(  \sigma
_{n}^{+}\sigma_{n+1}^{-}\right)  $, which commutes with $XX0$ Hamiltonian.

Our asymptotic results are not limited to the ASEP and they are also not
limited to conditioning on the current. One would obtain the same spatio-temporal
organization of the particle system in the limit of large activity, i.e., counting
the number of jumps irrespective of direction. In that case the modified rate
matrix (\ref{ModifiedRateMatrix}) would take the form $\widehat{W}
(s)=e^{s}(W_{+} +W_{-}) +W_{0}$. In this case the limit $\mbox{\rm
e}^{-s}\to0$ would directly lead to the $XX0$ Hamiltonian (\ref{HXX0}).

Similarly one can study non-conservative exclusion processes where particles
can be created or annihilated, see e.g. the 16-parameter family of reaction
diffusion processes defined in \cite{Alca93,Schu95}. This includes
diffusion-limited annihilation, the contact process and other well-studied
processes. For all these processes the quasi-stationary distribution for large
current or activity is the same as for the ASEP computed here. Hence our
asymptotic results apply to a large family of interacting particle systems.

The connection to the Circular Unitary Ensemble for $s\to\infty$ and the 
already known connections of the ASEP with the Gaussian Unitary 
Ensemble raise the question of further connections with random matrices. 
In particular, it would be interesting to perform the first corrections in 
$\rme^{-s}$ of the effective probability $P^{\eff}$, in order to check if it 
corresponds to the corrections of some potential on the unitary group 
$\mathcal{U}(N)$.

\section*{Acknowledgements}

V.P. thanks M. Salerno for stimulating discussions. This work was supported
by Deutsche Forschungsgemeinschaft.

\end{document}